\documentclass[aps,prb,a4paper,10pt,twocolumn,showpacs,floatfix,superscriptaddress,amsmath,amsfonts,amssymb,preprintnumbers,longbibliography]{revtex4-1}
\usepackage[T1]{fontenc}
\usepackage[utf8]{inputenc}
\usepackage{float}
\DeclareUnicodeCharacter{2212}{-}

\usepackage{dcolumn,graphicx,color,booktabs,microtype,afterpage}
\graphicspath{{./}{figure/}}
\usepackage[charter,greekuppercase=italicized]{mathdesign}
\usepackage{sidecap}
\renewcommand{\tablename}{Table}
\makeatletter\renewcommand{\fnum@figure}[1]{\figurename~\thefigure.~}\makeatother
\makeatletter\renewcommand{\fnum@table}[1]{\tablename~\thetable.}\makeatother

\newcount\hh \newcount\mm
\hh=\time \divide\hh by 60
\mm=\hh \multiply\mm by 60 \mm=-\mm
\advance\mm by \time
\def\now{\number\hh:\ifnum\mm<10{}0\fi\number\mm}

\usepackage[colorlinks,plainpages=false,linkcolor=blue,urlcolor=blue,citecolor=blue,pdfpagemode=UseNone,pdfstartview=FitBH]{hyperref}

\usepackage{nicefrac}

\newcommand{\ZB}{ZrB$_2$}
\newcommand{\ZVB}{Zr$_{0.96}$\-V$_{0.04}$\-B$_2$}
\newcommand{\HB}{HfB$_2$}
\newcommand{\HVB}{Hf$_{0.97}$\-V$_{0.03}$\-B$_2$}

\begin{document}

\makeatletter\renewcommand{\ps@plain}{%
\def\@evenhead{\hfill\itshape\rightmark}%
\def\@oddhead{\itshape\leftmark\hfill}%
\renewcommand{\@evenfoot}{\hfill\small{--~\thepage~--}\hfill}%
\renewcommand{\@oddfoot}{\hfill\small{--~\thepage~--}\hfill}%
}\makeatother\pagestyle{plain}


\title{Doping-induced superconductivity of ZrB$_2$ and HfB$_2$}

\author{N.\ Barbero}\email[Corresponding author: \vspace{8pt}]{nbarbero@phys.ethz.ch}
\affiliation{Laboratorium f\"ur Festk\"orperphysik, ETH Z\"urich, CH-8093 Zurich, Switzerland}

\author{T.\ Shiroka}
\affiliation{Laboratorium f\"ur Festk\"orperphysik, ETH Z\"urich, CH-8093 Zurich, Switzerland}
\affiliation{Paul Scherrer Institut, CH-5232 Villigen PSI, Switzerland}

\author{B.\ Delley}
\affiliation{Paul Scherrer Institut, CH-5232 Villigen PSI, Switzerland}

\author{T.\ Grant}
\affiliation{Escola de Engenharia de Lorena, Universidade de S$\tilde{a}$o Paulo, P.O. Box 116, Lorena, SP, Brazil}

\author{A.\,J.\,S.\ Machado}
\affiliation{Escola de Engenharia de Lorena, Universidade de S$\tilde{a}$o Paulo, P.O. Box 116, Lorena, SP, Brazil}

\author{Z.\ Fisk}
\affiliation{Department of Physics and Astronomy, University of California at Irvine, CA-92697, USA}

\author{H.-R.\ Ott}
\affiliation{Laboratorium f\"ur Festk\"orperphysik, ETH Z\"urich, CH-8093 Zurich, Switzerland}
\affiliation{Paul Scherrer Institut, CH-5232 Villigen PSI, Switzerland}

\author{J.\ Mesot}
\affiliation{Laboratorium f\"ur Festk\"orperphysik, ETH Z\"urich, CH-8093 Zurich, Switzerland}
\affiliation{Paul Scherrer Institut, CH-5232 Villigen PSI, Switzerland}

\begin{abstract}
\noindent
Unlike the widely studied $s$-type two-gap superconductor MgB$_2$, the chemically 
similar compounds ZrB$_2$ and HfB$_2$ do not superconduct above 1\,K. Yet, it has been shown 
that small amounts of self- or extrinsic doping (in particular with vanadium), can 
induce superconductivity in these materials. Based on results of different macro- and microscopic 
measurements, including magnetometry, nuclear magnetic resonance (NMR), 
resistivity, and muon-spin rotation ($\mu^+$SR), we present a comparative study 
of \ZVB\ and \HVB. Their key magnetic and superconducting features are 
determined and the results are considered within the theoretical framework 
of multiband superconductivity proposed for MgB$_2$. Detailed Fermi surface (FS) and
electronic structure calculations reveal the difference between MgB$_2$ and transition-metal
diborides.
\end{abstract}

\pacs{74.20.Fg, 74.25.−q, 75.40.Cx, 67.80.dk, 76.60.Cq}

\keywords{Unconventional supeconductivity, optimal doping, multiband effects, magnetism, nuclear magnetic resonance}

\maketitle\enlargethispage{3pt}

\vspace{-5pt}
\section{Introduction}\enlargethispage{8pt}
Borides, carbides, and nitrides were among the early compound 
superconductors discovered in the first half of the previous century\cite{Meissner1930} 
(see, e.g., Ref.\ \onlinecite{Hott2016} for a recent review).
However, it was only in 2001, with the 
discovery of superconductivity in MgB$_2$ at 39\,K,\cite{Akimitsu2001} 
that researchers intensified the search for superconductivity in 
other diborides. Based on a large number of studies, MgB$_2$ was identified as a two-band
two-gap superconductor. Its peculiar Fermi surface exhibits two-dimensional 
hole-like cylinders from the $p_{x,y}$ bands, a hole-like tubular network 
due to the bonding $p_z$ bands, and an electron-like tubular network 
due to the antibonding $p_z$ bands.\cite{Kortus2001} Due to this electronic configuration and to a distinct anisotropy of the electron-phonon
interaction strength, the electronic excitation spectrum of MgB$_2$ adopts two gaps in the superconducting phase: a 
large gap of 7.2\,meV in the $p_{x,y}$ ($\sigma$) bands, and a small gap 
of 2.8\,meV in the bonding and antibonding $p_z$ ($\pi$) bands. At the same 
time, an upper critical field anisotropy has been observed, with 
${\mu_0H_{c2}}^{c} \sim 30$\,T and ${\mu_0H_{c2}}^{ab} \sim 3$\,T at zero
temperature.\cite{Sologubenko2002}

Besides alkaline-earth metals (such as Mg), diborides of other elements have 
been proposed to be checked for superconductivity. 
Because of the presence of partially filled 3$d$, 4$d$, and 5$d$ orbitals, 
considered as promising for superconductivity, these new attempts involved 
icosagens (Al) and various $d$-type transition metals (T). The latter (TB$_2$), 
which are claimed to combine average coupling constants with comparable phonon
frequencies to MgB$_2$\cite{Heid2003} (due to the presence of light boron atoms),
were the natural candidates in this search. Unfortunately, these renewed efforts
proved unsuccessful and to date there are no reports of superconductivity for the
majority of TB$_2$ materials.

\ZB\ and \HB\ are two such non-superconducting refractory materials
with melting points of $\sim 3000$\,K, behaving essentially as Pauli 
paramagnets down to low temperatures. The electrons in the 4$d$ shell of zirconium (Zr) and those in 
the 5$d$ shell of hafnium (Hf) are less localized than those of the 3$d$ row.
In a recent study,\cite{Renosto2013} it was found that by replacing small amounts of Zr or Hf with V, 
the resulting compounds Zr$_{1-x}$V$_{x}$B$_2$ and Hf$_{1-x}$V$_{x}$B$_2$ 
are superconductors. Maximum superconducting temperatures 
$T_c = 8.33$\,K and 7.31\,K were reached in \ZVB\ and \HVB, respectively, at the
upper solubility limit of V ($x \approx 0.04$). X-ray powder diffraction (XRD) patterns\cite{Renosto2013}
indicate that an increase in V doping does not change the in-plane lattice parameter $a$, while it reduces 
the inter-layer distance $c$. At the same time, it has been shown that the structural and electronic properties of these compounds are
influenced by the presence of B vacancies.\cite{Dahlqvist2015} 

In Ref.\ \onlinecite{Renosto2013}, the properties of superconducting \ZVB\ were investigated by means of macroscopic
techniques.\cite{Renosto2013} In this work we aimed at combining macro- and 
microscopic techniques (including magnetometry, NMR, resistivity, and preliminary 
$\mu^+$SR experiments) on Zr-based diborides and extend our study to include 
the Hf-based compound. Since we succeeded in synthesizing 
samples with less magnetic impurities (below 10 ppm)
 with respect to the previous ones,\cite{Renosto2013} 
whenever relevant, a comparison between the two batches is included.
Our extensive data sets
allowed us to unravel clear analogies 
and differences between the T-diborides \ZVB\ and \HVB\ 
and the well-known MgB$_2$. In Sec.~\ref{ssec:NMR} we show that spin-lattice relaxation processes
in T-diborides are two orders of magnitude slower that in MgB$_2$, indicating significantly different
electronic structures.
Combined magnetometry (Sec.~\ref{ssec:SQUID}) and resistivity (Sec.~\ref{ssec:res}) 
measurements were performed to evaluate the upper and lower critical fields, respectively. These compounds
prove to be extreme type-II superconductors, as reflected by the high values ($\sim$ 100) of the Ginzburg-Landau $\kappa = \nicefrac{\lambda}{\xi}$ parameter. The London penetration depth $\lambda$ was evaluated through $\mu^+$SR (Sec.~\ref{ssec:musr}) experiments and the coherence length $\xi$ via upper critical field measurements (Sec.~\ref{ssec:res}). Thanks to the efficient complementarity of these four techniques, we argue that, besides the qualitatively different Fermi surfaces of MgB$_2$ and transition-metal diborides, in both cases we are dealing with $s$-wave superconductors. While the rather high $T_c$ of MgB$_2$ is understood as a consequence of a favorable electronic structure and electron-phonon interaction, the drastic effect of V-doping in the T-diborides is still
rather surprising.

\begin{figure}[tbh]
  \centering
  \includegraphics[width=0.33\textwidth]{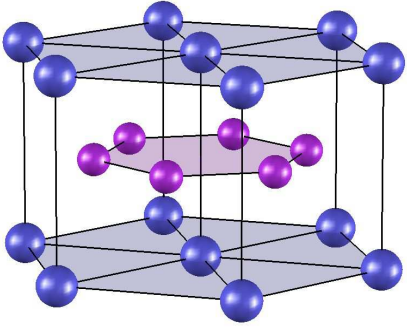}
  \caption{\label{fig:ZrB2_structure}AlB$_2$ structure of the transition-metal diborides highlighting the graphene-like 
  B layers (magenta atoms) and the hexagonal metal layers (blue atoms).}
\end{figure}

\vspace{-5pt}
\section{Electronic structures of Z\lowercase{r}B$_2$ and H\lowercase{f}B$_2$\label{sec:fermisurf}}\enlargethispage{8pt}

Both pure and V-doped compounds crystallize in the layered AlB$_2$ structure with a $P6/mmm$ hexagonal
space group, where the Zr (or Hf) atoms and the B atoms occupy, respectively,
the $1a$ (0,0,0) and $2d$ (1/3, 2/3, 1/2) positions.
As in MgB$_2$, their crystal lattices are characterized by hexagonal metal
layers alternating with graphite-like B layers (see Fig.~\ref{fig:ZrB2_structure}).

\begin{figure}[tbh]
  \centering
  \includegraphics[width=0.35\textwidth]{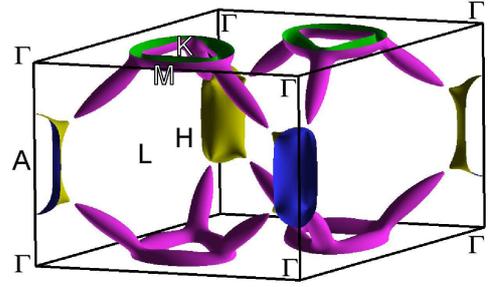}  
  \caption{\label{fig:FS_ZB} Fermi surface of \ZB.}
\end{figure}

Previous results of band-structure calculations\cite{Shein2008,Grechnev2009}  and Fermi surface (FS) representations\cite{Shein2008} are available in the literature for both \ZB\ and \HB. Since, however, the printed version of the published FS of \ZB\ suffers from low quality, we chose to present the result of our own calculation in Fig.~\ref{fig:FS_ZB}, intended to serve for a comparison with the published FS of Mg$B_2$,\cite{Mazin2003} well known for its amazingly high critical temperature $T_c$ of the order of 40\,K for the onset of superconductivity. Our density-functional calculations use an LDA approach\cite{Perdew1992} based on the DMol$^3$ band-structure modeling program.\cite{Delley1990,Delley2000} The self-consistent field (SCF) was accomplished using  the standard DNP\cite{Delley1990} 
variational basis set and a $\Gamma$-centered $12\times 12\times 12$ $k$-mesh in the reciprocal cell. Pseudized scalar relativistic corrections\cite{Delley1998} were applied. The calculations are based on the experimental geometries published in Ref.~\onlinecite{Bsenko1974}.
 The FS plot was generated using the XCrySDen program\cite{Kokalj1999} based on the DMol$^3$ output for a $99\times 99\times 99$ $k$-mesh.

The \ZB\ FS consists of four short cylindrical hole-type pockets around the $A$ point and four tripod-shaped electron-type pockets, each consisting of a triangular ring around the $K$ point and three elliptic extensions near the $\mathrm{\Gamma}AH$ plane. Since the FS of \HB\ exhibits essentially the same features, we refrained from presenting it in a separate figure. These features are really quite different from those of the FS of MgB$_2$. Most obvious is the reduction of the two almost two-dimensional hole sheets centered near the $A$ point for MgB$_2$ to one much smaller and 3-D type pocket for Zr- and Hf-diboride. Likewise the 3-D parts are quite different in shape for MgB$_2$. It is therefore not surprising that the two borides investigated here are less favorably conditioned for superconductivity than MgB$_2$ and indeed for both pure compounds, no superconductivity was detected above 1\,K. The band-structure calculations for the T-borides considered here indicate that in both cases the density of electronic states  at the Fermi energy $N(E_\mathrm{F})$ is located in a region where $N(E)$ exhibits a pseudo-gap, i.e., is much reduced. It is thus remarkable that a very small (4--5\%) V-for-Zr or V-for-Hf substitution results in onsets of superconductivity up to approximately 8\,K. More details on the crystal structure and defect-induced phase instabilities of these types of compounds 
are discussed in Ref.~\onlinecite{Renosto2013}.

%


\vspace{-5pt}
\section{Experimental details\label{sec:details}}\enlargethispage{8pt}

Polycrystals of \ZB, \HB, \ZVB, and \HVB\ were synthesized via boron carbide
reduction and structurally characterized as described in Ref.\ \onlinecite{Renosto2013}. 
For our magnetic measurements, the samples in form of fine powders, with masses between
50 and 100\,mg, were sealed in Teflon (PTFE) tubes. The NMR investigations including line-shape and spin-lattice relaxation measurements were performed in an applied magnetic field of 3.505\,T, since higher fields would reduce the possibility to resolve 
the quadrupolar effects. In our case, the $^{11}$B nucleus (spin 
$I=\nicefrac{3}{2}$) proved to be the most suitable one, since it allows a direct comparison 
between the four samples and it is four times more abundant than $^{10}$B. 
The NMR signals were monitored by means of standard spin-echo sequences, consisting in 
${\pi}/{2}$ and $\pi$ pulses of 2 and 4\,$\mu$s, respectively, with recycle delays 
ranging from 10 to 100\,s, depending on the temperature ranging between 4 and 295\,K. 
The NMR line-shapes were obtained via the fast Fourier Transform (FFT) of the echo 
signal which, due to the high sensitivity of  $^{11}$B, could be acquired using relatively few scans (from 4 to 2048).
The spin-lattice relaxation times $T_1$ were measured with the inversion recovery method,
using a  $\pi$-${\pi}/{2}$-$\pi$ pulse sequence. The magnetometry measurements were performed by using a commercial 
Magnetic Property Measurement System (MPMS XL) from Quantum Design, 
equipped with a 7-T magnet and covering the temperature range from 2 to 400\,K. For the resistivity
 measurements, the samples were densely packed in 
cylindrical pellets with a diameter of 1.4\,mm and a thickness of 5 mm, 
produced by applying high external pressures. The electrical contacts with a four-probe
configuration were made by means of an electrically-conducting silver epoxy.

Preliminary $\mu^+$SR measurements were made at the GPS spectrometer 
of the S$\mu$S facility of Paul Scherrer Institute (PSI) in Villigen, Switzerland. The available 
sample mass (300\,mg) was sufficient to stop the 4-MeV muons without additional
degraders and with a minimal background signal. To avoid pinning effects, 
well known for distorting the vortex lattice (VL) in the superconducting phase of 
MgB$_2$,\cite{Niedermayer2002} the transverse-field (TF) muon-spin rotation measurements
were made at the highest field available (0.6\,T).

\vspace{-5pt}
\section{Experimental results and discussion\label{sec:results}}\enlargethispage{8pt}
\subsection{Nuclear Magnetic Resonance\label{ssec:NMR}}

\begin{figure}[tbh]
  \centering
  \includegraphics[width=0.45\textwidth]{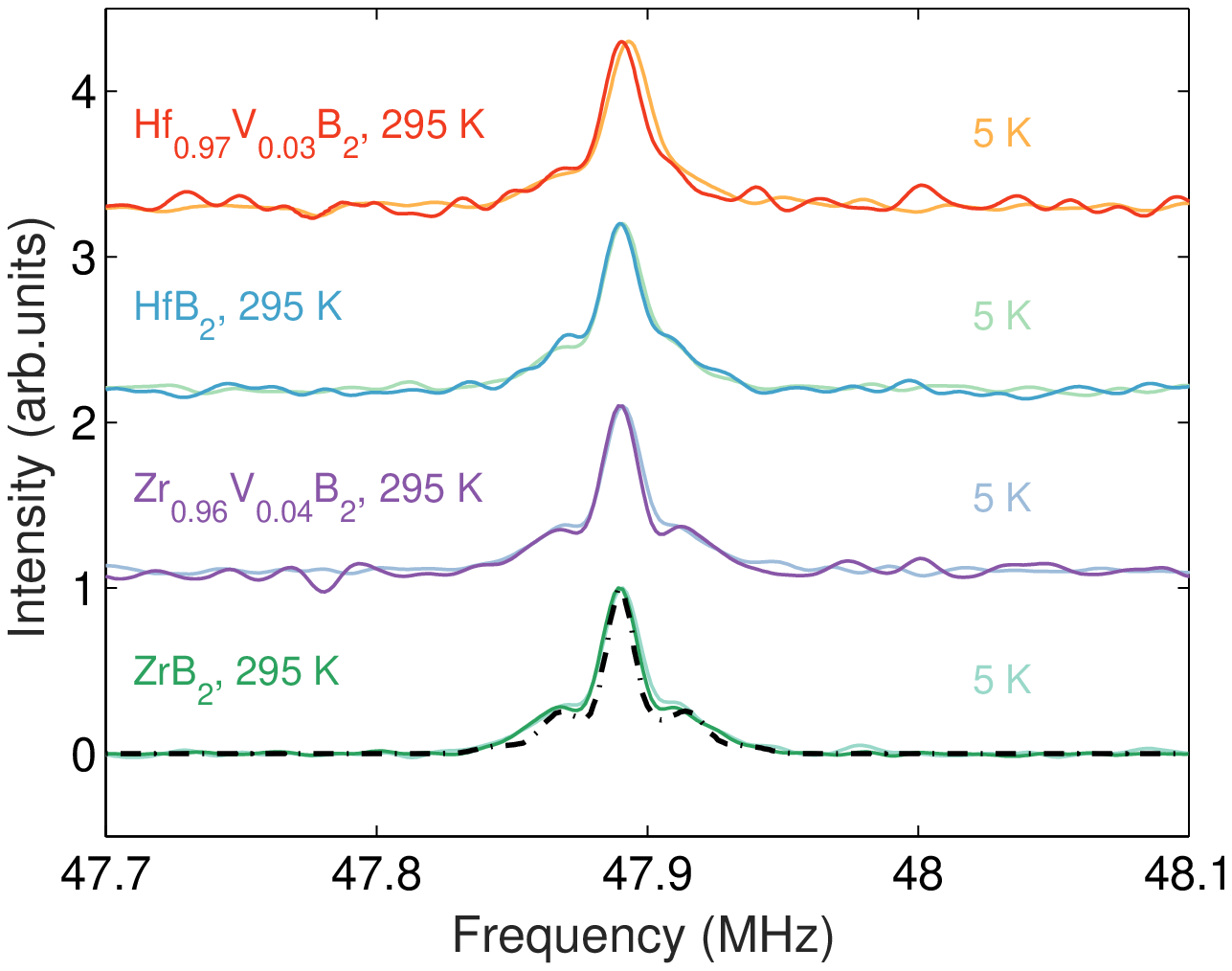}
  \caption{\label{fig:Diborides_NMR_lines_comparison} ${}^{11}$B NMR lines in the four TB$_2$ compounds, respectively - from bottom to top, \ZB, \ZVB, \HB, and \HVB. The measurements were performed at 3.505\,T and for each sample the lines measured at 5 and at 295\,K are superposed, showing no relevant change. The dashed black line superposed on the \ZB\ spectra represents a simulation of a S = $\nicefrac{3}{2}$ powder pattern with small quadrupole splitting (reflected by the two shoulders around the central transition).}
\end{figure}

\begin{figure}[tbh]
  \centering
  \includegraphics[width=0.45\textwidth]{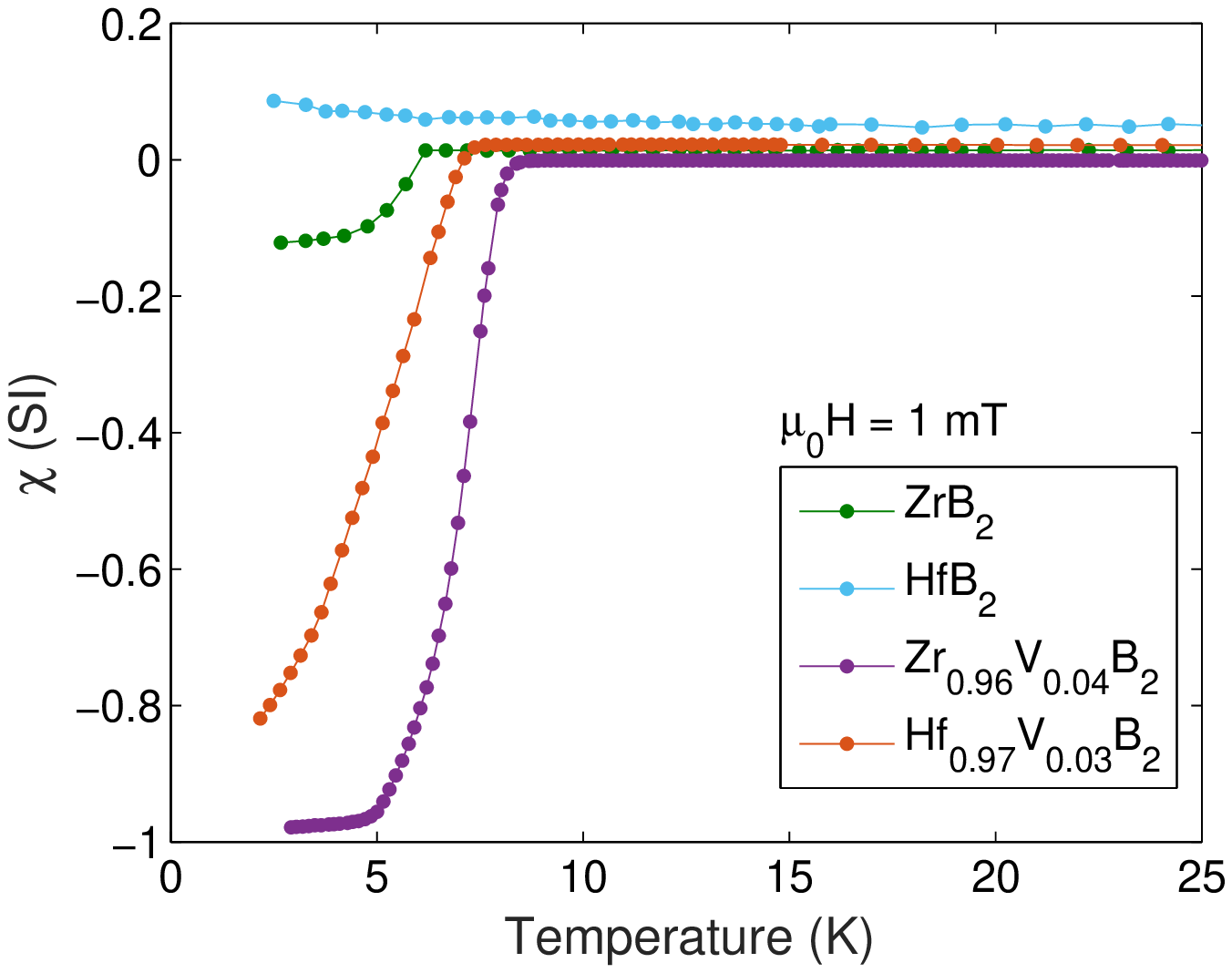}
  \caption{\label{fig:chi_vs_T_10Oe_allsamples} Magnetic susceptibility data [$\chi(T)$ in SI units] 
at 1\,mT for \ZB, \HB, \ZVB, and \HVB. The undoped samples (and the doped ones above $T_c$) exhibit
diamagnetic behavior (plateaux). Small ZrB$_{12}$ impurities in \ZB\ induce a superconducting transition at $\sim 6$\,K. \ZVB\ and \HVB\ become superconductors at $T_c =  8.33$\,K and 7.31\,K, respectively.}
\end{figure}

The ${}^{11}$B NMR lines of all the samples (both pure and V-doped) 
were measured from 5 to 295\,K; typical data are shown in Fig.~\ref{fig:Diborides_NMR_lines_comparison}. 
The reference ${}^{11}$B NMR frequency in an applied magnetic field of  3.505\,T was 
evaluated to be $\nu_0 = 47.8844$\,MHz. 
In our case, the NMR lines exhibit peaks which are very close to the reference, with an absolute positive shift
of only about 6\,kHz, corresponding to a Knight shift of 120\,ppm. 
In the covered temperature range between 5 and 295\,K the ${}^{11}$B NMR lines practically 
coincide (see Fig.~\ref{fig:Diborides_NMR_lines_comparison}), implying temperature-independent
Knight shifts for all the measured samples, compatible with the $\chi(T)$ plateaux
observed in the magnetometry data, measured under zero-field cooled 
conditions (see Fig.~\ref{fig:chi_vs_T_10Oe_allsamples}).
The trend of the Knight shift in the superconducting phase could not be resolved because of the appreciable
width of the resonance signal.
As the line position, also the full width at half maximum (FWHM) is practically constant upon varying the 
temperature. For the Zr-based samples its value is 13\,kHz (with 1\,kHz of additional broadening below $T_c$), 
while for those containing Hf the width is 14\,kHz (+1\,kHz at low temperatures). The typical quadrupole splitting 
$\nu_Q$ in \ZB\ and \HB\ is approximately the same, i.e. 54\,kHz. This relatively small value implies 
a rather small electric-field gradient (EFG), especially if compared with MgB$_2$, for which 
$\nu_Q \simeq 860$ kHz,\cite{Papavassiliou2001} which is a signature of a different electronic charge distribution,
confirmed by the different orbitals involved in the bonds, i.e. only $s$ and $p$ orbitals for MgB$_2$ and also $d$ orbitals for T-borides.

For typical powder spectra of a $I = \nicefrac{3}{2}$ nucleus with a small quadrupole splitting, 
an analytical expression for the line-shape can be derived by considering the quadrupole term as a first-order 
perturbation in the main Zeeman Hamiltonian.\cite{Goc1987, Cohen1954} As shown in Fig.~\ref{fig:Diborides_NMR_lines_comparison}, 
the experimental spectrum and the simulated line-shape (dashed line) for \ZB\ agree quite well.
The simulation of the powder spectrum, employing a Matlab code, performs the integration according to Euler's method over all
the possible orientations of the NMR line-shape factor\cite{Goc1987} $F$, i.e., the integration of the transition frequencies, derived from the quadrupolar theory and weighted using a Gaussian broadening function.\cite{Alderman1986, Hodgkinson2000}
From a quantitative analysis, the electric-field gradient (EFG) tensor can be evaluated using

\begin{equation} \label{eq:EFG}
eq = \frac{2I(2I-1)\,h\nu_Q}{3eQ},
\end{equation}
with $eq$ the largest EFG component (parallel to the applied magnetic field), 
$I$ the nuclear spin, and $Q$ the quadrupole moment of the nucleus. 
By considering the hexagonal symmetry of the AlB$_2$ structure, 
we can assume that the in-plane anisotropy parameter $\eta = (V_{xx}-V_{yy})/V_{zz} = 0$ and, therefore,
$V_{xx} = V_{yy}$. Since the $V_{i,j}$ tensor is traceless, by evaluating $eq = V_{zz}$ from the simulated $\nu_Q$ 
data (see Eq.~\ref{eq:EFG}), we get $V_{zz} = 1.1\times10^{20}$\,Vm$^{-2}$ and $V_{xx} = V_{yy} = -5.5\times10^{19}$\,Vm$^{-2}$.

\begin{figure}[tbh]
  \centering
  \includegraphics[width=0.45\textwidth]{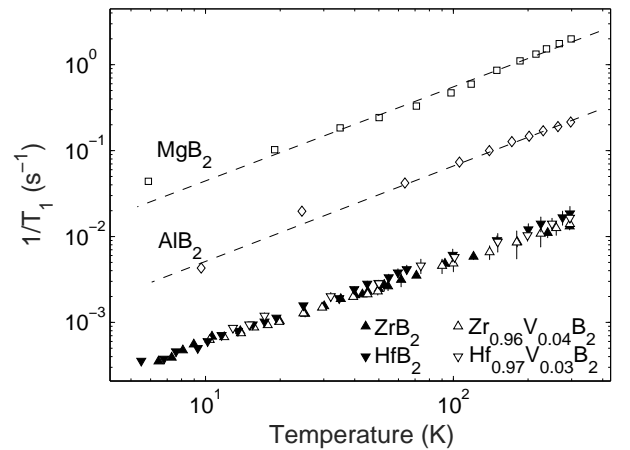}
  \caption{\label{fig:Diborides_T1_compar} ${}^{11}$B NMR $T_1^{-1}(T)$ data 
at 3.505\,T for all the investigated samples, compared with MgB$_2$ and AlB$_2$ data
(from Ref.\ \onlinecite{Baek2002}). From a standard fit,
assuming only one spin-lattice relaxation time,\cite{Mcdowell1995}
we obtain a linear trend, with Korringa constants as reported in Table~\ref{tab:summary}. 
Materials with transition metal cations, such as Zr and Hf, exhibit slow spin-lattice relaxation processes,
two orders of magnitude slower than in MgB$_2$, whereas AlB$_2$ is an intermediate case.}
\end{figure}

As summarized in Table~\ref{tab:summary}, the Knight shift values are of the same order of magnitude as in MgB$_2$,\cite{Kotegawa2001,Papavassiliou2001,Baek2002,Pavarini2003} 
but we note a difference of two orders of magnitude (!) in 
the spin-lattice relaxation rates. The measured $T_1T$ value for MgB$_2$  is $1.8\times 10 ^2$\,sK which, 
by considering its 70\,ppm Knight shift, implies an experimental Korringa constant $S_0 \equiv T_1TK^2 = 8.85\times 10^{-7}$\,sK, 
approximately 3 times smaller than the theoretical value $S_{\mathrm{th}} = \hbar(\gamma_e/\gamma_{\mathrm{B}})^2 /(4\pi k_{\mathrm{B}}) = 2.56 \times 10^{-6}$\,sK. Upon V-doping, the very slow relaxation processes in T-diborides are non significantly
altered but, nevertheless, this apparently insignificant doping induces superconductivity at relatively high critical temperatures. Furthermore, due to strong covalent bonds, the largest contribution to the electronic density of states (DOS) at the Fermi level  $D(E_F)$ is due to itinerant electrons in the boron layers. With respect to AlB$_2$ and even more so to MgB$_2$, $D(E_F)$ of our compounds is drastically reduced, as clearly confirmed by the spin-lattice relaxation data, in turn in good agreement with our theoretical calculations on the electronic structure of \ZB\ and \HB.

\begin{table}
\centering 
 \caption{\label{tab:summary} Key NMR parameters for the investigated samples compared with those of AlB$_2$ and MgB$_2$.}
  \begin{ruledtabular}
  \begin{tabular}{ l  c  c  c } 
    Material & Shift (ppm)  & $T_1T$ (10$^4$ s\,K) & $S_0$ (10$^{-4}$ s\,K) \\ \hline
    ZrB$_2$  & 120 &  1.81 $\pm$ 0.07 & 2.60 $\pm$ 0.18 \\ 
    Zr$_{0.96}$V$_{0.04}$B$_2$ & 120 &  1.89 $\pm$ 0.07 & 2.72 $\pm$ 0.11 \\ 
    HfB$_2$ & 140  &  1.67 $\pm$ 0.05 & 3.27 $\pm$ 0.12 \\ 
    Hf$_{0.97}$V$_{0.03}$B$_2$  & 140 & 1.56 $\pm$ 0.07 & 3.06 $\pm$ 0.09  \\
    AlB$_2$  & -10 & 0.14 $\pm$ 0.04 & 0.009 $\pm$ 0.001  \\
    MgB$_2$  & 70 &  0.018 $\pm$ 0.006 & 0.009 $\pm$ 0.001 \\ 
  \end{tabular}
 \end{ruledtabular}
\end{table}

The raw data, i.e., ${}^{11}$B NMR $T_1$ inverse saturation recovery curves, for the temperatures above $T_c$ were fitted by assuming the standard magnetization recovery formula for a single spin-lattice relaxation time.\cite{Mcdowell1995} However, due to the expected high anisotropy of $\mu_0H_{c2}$, and by analogy with MgB$_2$,\cite{Sologubenko2002} in powder samples we expect grains with different orientations, i.e., 
where the applied magnetic field lies in the $ab$ plane or is parallel to the $c$ axis. \textit{A priori} the orthogonal and parallel magnetic susceptibilities and the upper critical field depend on the direction of the field. 
The existence of nonequivalent grains, due to the anisotropy of the upper critical field, is also confirmed by magnetometry (Fig.~\ref{fig:chi_vs_T_10Oe_allsamples}) and resistivity (Fig.~\ref{fig:Diborides_R_vs_T_all}) measurements, which show a superconducting width transition $\Delta T \sim 2.5$\,K, hence suggesting the persistence of inhomogenous domains. In a first approximation and following a procedure employed in the MgB$_2$ case,\cite{Baek2002, Pavarini2003}, we can fit the data, by assuming two relaxation processes (see Eq.~\ref{eq:IRfit}), related to the normal phase and to the superconducting phase, respectively:
\begin{equation} \label{eq:IRfit}
I(\tau_d) \propto \alpha \exp(\tau_d/T_{1s})^\beta + 
(1 - \alpha) \exp(\tau_d/T_{1n})^\beta.
\end{equation}
Here $\tau_d$ is the time delay in the NMR pulse sequence, $T_{1s}$ and $T_{1n}$ the spin-lattice relaxation times
of the superconducting and normal grains, respectively, $\alpha$ the superconducting volume fraction and $\beta$ 
the stretching parameter (close to 1 in this case). We assume that $\alpha$ is a temperature-dependent 
fit parameter, ranging from $\sim 1$ (in case of maximum superconducting fraction, 
as evaluated from magnetometry data), down to 0. In the transition region the two plateaux are connected with a sigmoidal function.

\begin{figure}[tbh]
  \centering
  \includegraphics[width=0.45\textwidth]{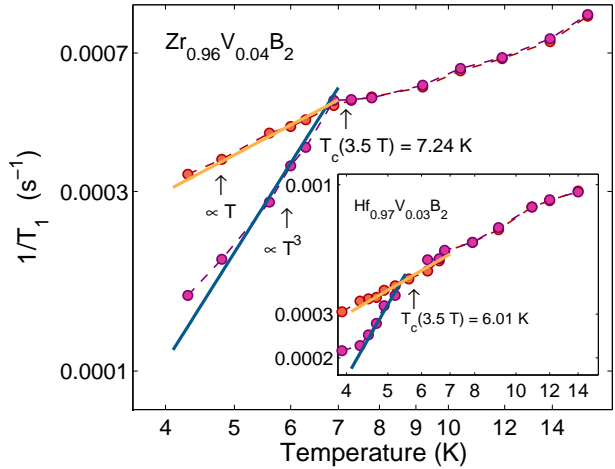}
  \caption{\label{fig:diborides_gap} ${}^{11}$B NMR $1/T_1(T)$ data at 3.505\,T for \ZVB\ (main plot) and \HVB\ (inset), representing
  the two relaxation times of Eq.~\ref{eq:IRfit}, which result from fitting the inversion recovery curves.
  Both superconducting samples exhibit a conventional metallic trend (red points) and a superconducting dropdown (magenta points); The blue lines are power laws with critical exponent 3, that would be a signature of $d$-wave superconductivity. In both the samples we argue that the faster relaxation rates below 5.5\,K with respect to the power-law support the hypothesis of $s$-wave pairing.}
\end{figure}

The $T_1^{-1}(T)$ spin-lattice relaxation data above $T_c$ follow the linear behavior in $T$ of a simple metal (see Fig.~\ref{fig:Diborides_T1_compar}), where nuclear relaxation occurs mostly via interactions with the conduction electrons. 
On the other hand, below $T_c$, the superconducting grains may exhibit two trends: a power-law with an integer exponent, typical of anisotropic superconductors, or an exponential trend, as expected for $s$-wave superconductors. If the sample is not perfectly homogeneous, 
a second relaxation component could persist as a linear trend associated to normal grains, as shown in Fig.~\ref{fig:diborides_gap}.
The hypothesis of the two relaxation times is justified by the good fit results that we obtain for the evaluation of spin-lattice relaxation rates $R_1 = T_1^{-1}$.
A quantitative analysis is however hampered by a contribution to the relaxation from the flux vortex lines (i.e., their thermal motion).\cite{Jung2001, Rigamonti1998} Since we performed a field cooling (FC) measurement, the formation of a flux line lattice (FLL) with a regular arrangement of vortices, most likely with hexagonal symmetry, is expected. The geometric parameter of this lattice is the intervortex spacing $d(\mu_0H) = (2\Phi_0/3^{1/2}\mu_0H)^{1/2}$, which implies $d(3.505 T) = 26$ nm in our case. This value is approximately 6 times the diameter of the vortices $\xi$, as evaluated in Sec.~\ref{ssec:res}, implying that the measured 
T$_1$ values consist of the sum of a slow contribution from outside the vortex cores and a faster contribution from the normal region within the vortices.\cite{Rigamonti1998}

We speculate that the considerable anisotropy of the upper critical field of MgB$_2$,\cite{Sologubenko2002} is also a characteristic of our materials. A rigorous confirmation would be obtained by relevant experiments on single crystals as, e.g., thermal conductivity measurements. In general, $s$-wave superconductors exhibit an exponential decrease of $T_1^{-1}(T)$ well below $T_c$, from which  the gap value $\Delta$ can be extracted.
The appearance of a Hebel-Slichter coherence peak is usually interpreted as confirming the Cooper-pairing with spherical symmetry. In our case, the absence of a coherence peak does not rule out an $s$-wave parity, since the size of the peak can be significantly reduced by the pair-breaking mechanism in the presence of high fields.\cite{Masuda1969}
Due to the above-mentioned complexity of the relaxation processes and the quality of our data, it is impossible to extract the gap value $\Delta$ from the exponential decrease of the spin-lattice relaxation time. In any case a first evidence for $s$-wave superconductivity is the increasing deviation from a power-law behavior with an exponent 3 towards lower temperatures (see Fig.~\ref{fig:diborides_gap}).
To justify the similarities between the phonon-mediated $s$-wave superconductivity mechanism in MgB$_2$ and our V-doped samples we note that vanadium, given its $3d^3$ orbital, has one more electron, if compared to Zr ($4d^2$) and Hf ($5d^2$). 
An analogous electron doping is confirmed in the case of MgB$_2$, where the $s$-states of Mg are pushed up by the B $p_z$ orbitals and, therefore, fully donate their electrons to the boron-derived conduction band.\cite{Kortus2001} This doping mechanism occurs also in the opposite direction (reduction of the $T_c$ value) in MgB$_2$. In this case, both the substitution of Mg with Li (hole doping) and of boron with carbon or Al (electron doping) reduce the $T_c$ of the material.\cite{Karpinski2008} In this case, it is claimed that the electrons fill the $\sigma$ band and holes occupy the $\pi$ band, therefore making charge compensation impossible. Furthermore, a recent paper\cite{Gil2013} supports the hypothesis of two gaps in \ZVB\ from critical current density $J_c$ measurements in different fields. In fact, $J_c$ can suitably be fitted by the sum of two contributions $J_1$ and $J_2$, respectively, related to the first and the second gap, following an exponential trend as a function of the applied magnetic field.

\subsection{SQUID Magnetometry\label{ssec:SQUID}}

\begin{figure}[tbh]
  \centering
  \includegraphics[width=0.45\textwidth]{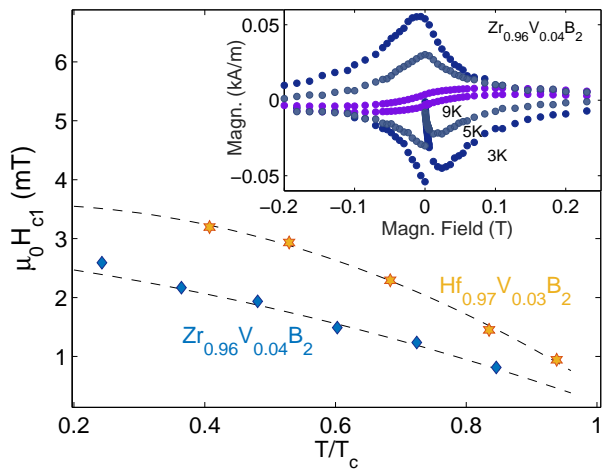}
  \caption{\label{fig:Hc1_trend} The temperature dependence of $\mu_0H_{c1}$ exhibits a negative curvature. By using
  parabolic fits (dashed lines) the approximate $H_{c1}(0)$ values are obtained: 2.5 mT for \ZVB\ and 3.6 mT for \HVB. Inset: the $M$ vs.\ $H$ plots (zero field cooling) in \ZVB\ at 3, 5, 9, and 12\,K}
\end{figure}

SQUID magnetometry measurements were made on all the samples (\ZB, \HB, \ZVB, and \HVB).
The high sensitivity (10$^{-10}$\,Am$^2$) of the Reciprocating Sample Option (RSO) of the MPMS magnetometer allowed us to detect small impurities. In particular, in the \ZB\ and \ZVB\ samples a small mass fraction of about 0.5\,\% exhibits a superconducting transition at 5.5 K, the typical $T_c$ of ZrB$_{12}$ impurities. Smaller impurity contributions are also visible from the $\chi(T)$ plots at fields between 0.1 and 7\,T, exhibiting a steady increase below 20\,K. 
Therefore, an accurate measurements of $T_c(H = 0)$ was achieved by applying 
a small magnetic field of 1\,mT. The obtained values of $T_c$ are 8.33\,K for \ZVB\ and 7.31\,K for \HVB.
As shown in the inset of Fig.~\ref{fig:Hc1_trend}, the typical type-II SC cycles can be observed below $T_c$ ;
above $T_c$, a clear diamagnetic trend is confirmed with a typical 
$\chi_m = -7.5\times 10^{-10}$\,m$^3$mol$^{-1}$, 
a value approximately 5 times smaller than bismuth.

The low-field region (from 0.2 to 6\,mT), exhibits an initial linear trend in $M(H)$. It is possible to extract an approximate value of $\mu_0H_{c1}$, defining it as the field at which the deviation from the linear trend (called \textit{Meissner line}) exceeds the sensitivity of the instrument. Performing this analysis, for each of the $M$ vs.\ $H$ curves, we obtain $\mu_0H_{c1}(T)$, as shown in Fig.~\ref{fig:Hc1_trend} for both \ZVB\ and \HVB. According to the two-band Ginzburg-Landau theory applied to MgB$_2$, the lower critical field exhibits a change in concavity (from negative to positive,   upon cooling) at $T/T_c \sim$ 0.5.\cite{Askerzade2002} Since this change is scarcely distinct and the model depends upon the interband mixing of the two order parameters and of their gradients, it is difficult to interpret the $\mu_0H_{c1}(T)$ trend which, as reported in the literature for MgB$_2$,\cite{Sharoni2001, Li2001} can be fitted even with a line. On the other hand, the previously reported \ZVB\ data,\cite{Renosto2013} 
show a pronounced upturn, which is not present in our case. The reason for this discrepancy is unclear at the moment.

\subsection{Resistivity\label{ssec:res}}
\begin{figure}[tbh]
  \centering
  \includegraphics[width=0.45\textwidth]{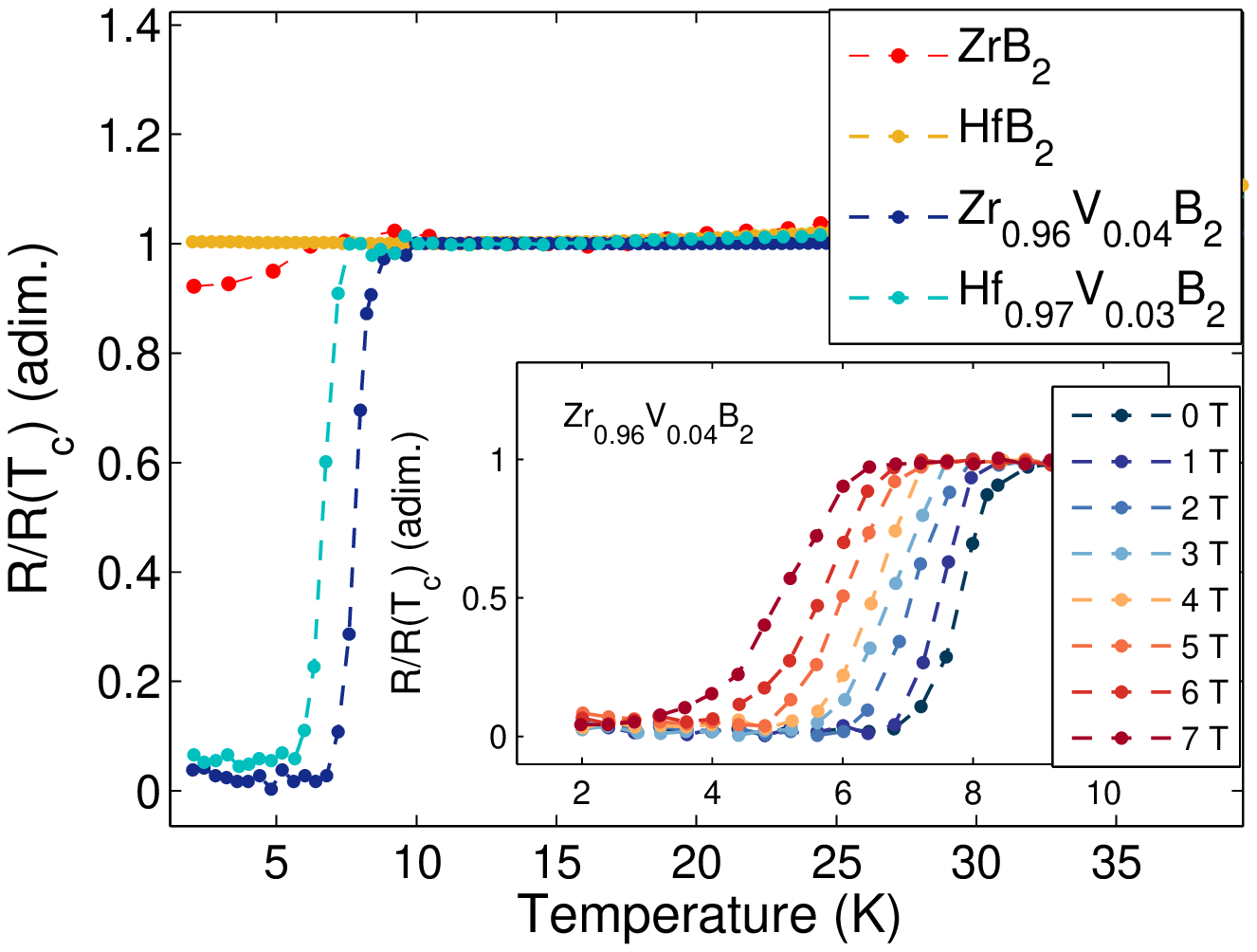}
  \caption{\label{fig:Diborides_R_vs_T_all}$R$ vs. $T$ data for all  the samples (\ZB, \HB, \ZVB, and \HVB) in zero field. The superconducting dropdown in \ZVB\ and \HVB\ are evident and confirm the SQUID magnetometry measurement. On the other hand, undoped samples exhibit a constant plateau, as 
  reflected also by the constance of Knight shift (see Sec.~\ref{ssec:NMR}). We attribute the small dropdown of \ZB\ to ZrB$_{12}$ impurities. The inset shows the resistivity data of \ZVB\ between 2 and 35\,K has been measured at different fields (from 0 to 7 T), confirming a reduced steepness towards higher fields ($\Delta T$ ranging from 2.5 to 3.5\,K towards higher fields) and the expected negative shift.}
\end{figure}

Systematic resistivity measurements were performed in zero field on all the samples
(\ZB, \HB, \ZVB, and \HVB), as shown in Fig.~\ref{fig:Diborides_R_vs_T_all}, 
and on the superconducting samples \ZVB\ (see the inset of Fig.~\ref{fig:Diborides_R_vs_T_all}) 
and \HVB\ in magnetic fields up to 7\,T. 

Each resistance measurements in zero field (ZF) was performed from 2 to 310\,K, while in field we focused our 
attention on the superconducting transition region (range from 2 to 10\,K). Due to the small resistivity values, i.e., $\rho(T_c) \sim$ 0.7 $\mu \Omega$cm for \ZVB\ and 0.8 $\mu \Omega$cm for \HVB, we argue that we can analyze our data within the clean limit approximation.

Based on the data at different fields, the $\mu_0H_{c2}(T)$ values were evaluated. According to the theory of Werthamer, Helfand, and Hohenberg (WHH),\cite{Helfand1966} in the clean limit\cite{Khim2011} and for small spin-orbit couplings,\cite{ Werthamer1966} we have

\begin{equation} \label{eq:WHH}
\mu_0H_{c2}(0) = -0.73\,T_c\,\left.\frac{dH_{c2}}{dT}\right|_{T=T_c}.
\end{equation}

Within this approximation, $\mu_0H_{c2}(T_c)$ can be fitted with a parabolic curve:

\begin{equation} \label{eq:WHHfit}
\mu_0H_{c2}(T) = \mu_0H_{c2}(0)\,[1-(T/T_c)^2]
\end{equation}
\begin{figure}[tbh]
  \centering
  \includegraphics[width=0.45\textwidth]{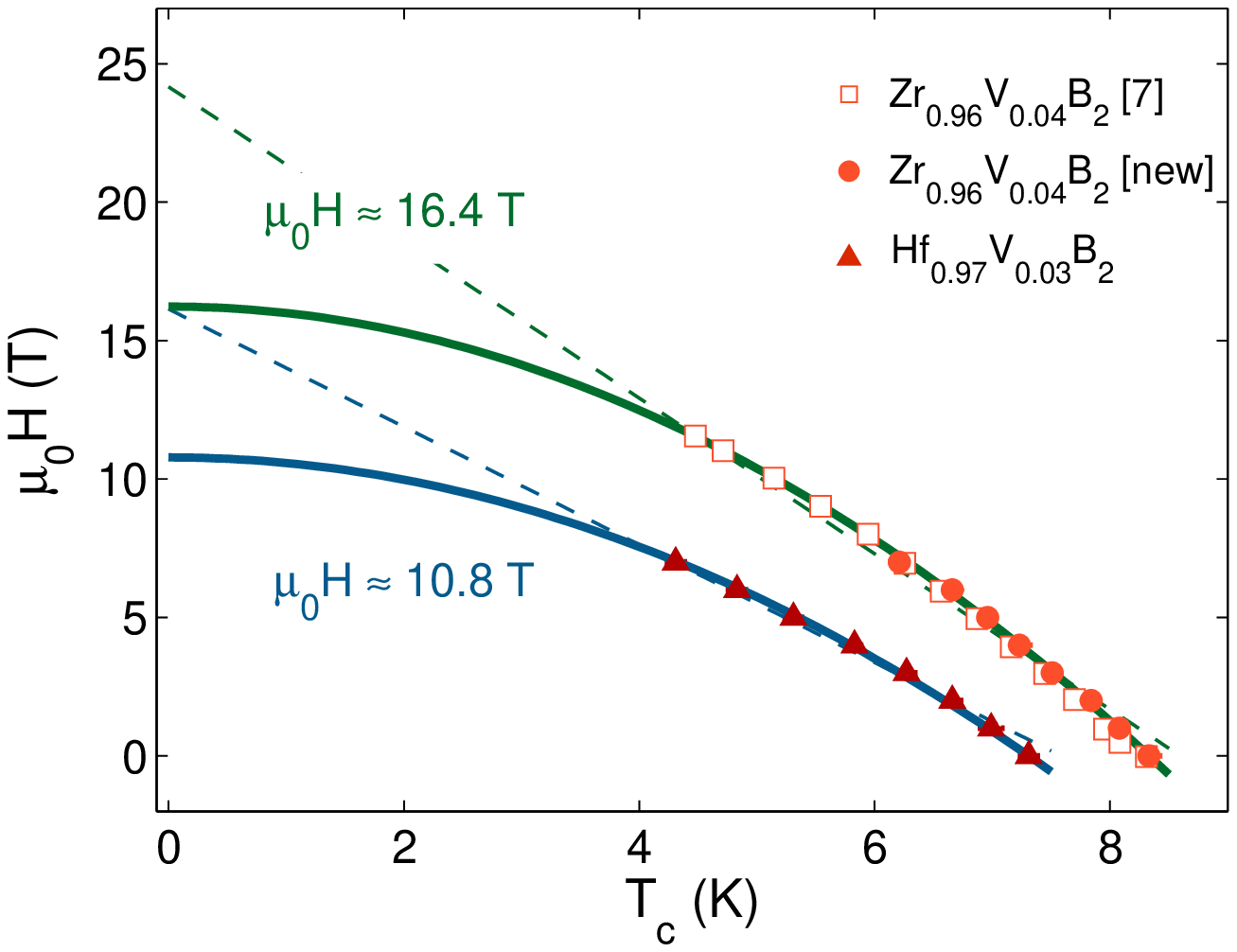}
  \caption{\label{fig:Diborides_Hc2_WHH_analyses} $\mu_0H_{c2}(T_c)$ plots for \ZVB\ (filled orange circles) and
  \HVB\ (filled red triangles) are derived from resistivity measurements at different field from 0 to 7\,T. Via the WHH model, that implies
  a parabolic fit (green and blue continuous lines), we can evaluate $\mu_0H_{c2}(0)$, i.e. 16.4\,T for \ZVB\ and 10.8\,T for \HVB. Since Hf is heavier than Zr, we argue that the lower value in the upper critical field is related to the lower average phonon frequency. The old data (Ref.\ \onlinecite{Renosto2013}, empty squares) and the new ones (circles) are in good agreement.}
\end{figure}

From the Ginzburg-Landau formula 
$\xi(0) = [\phi_0/(2 \pi \mu_0 H_{c2}(0))]^{1/2}$,
our estimate of the coherence lengths $\xi(0)$ in both \ZVB\ and \HVB\, are 4.5(1) and 5.5(1)\,nm, respectively.
By numerically solving

\begin{equation} \label{eq:Bc1Bc2}
\frac{\mu_0H_{c2}(0)}{\mu_0H_{c1}(0)} = \frac{2 \kappa^2}{\ln \kappa },
\end{equation}

we finally obtain the Ginzburg-Landau $\kappa$ parameter, with a value of 125 for \ZVB\ and
80 for \HVB, respectively indicating the strong type II nature of these superconductors. From
$\kappa = \lambda/\xi$, the London penetration depth $\lambda(0)$ in the two materials, is 570 and 445\,nm, respectively.

\subsection{Muon-spin rotation results in the SC phase\label{ssec:musr}}

\begin{figure}[bth]
  \centering
  \includegraphics[width=0.45\textwidth]{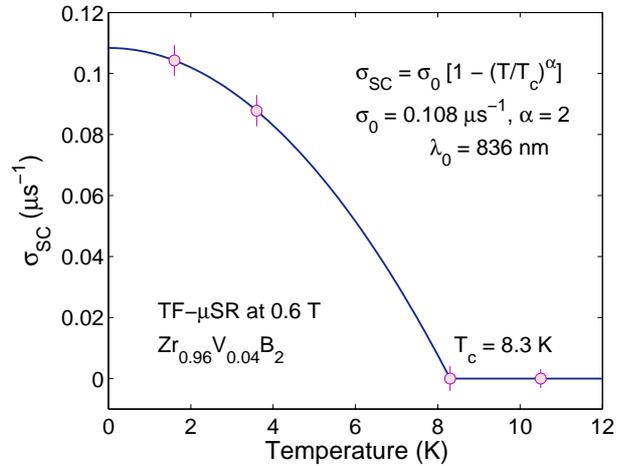}    
  \caption{\label{fig:ZrVB2_TF_MuSR} Transverse-field $\mu_SR$ data in the superconducting phase of  \ZVB\ in a magnetic 
  field $\mu_0 H = 0.6$\,T.}
\end{figure}

Values of similar magnitude for the magnetic field penetration depth were obtained from preliminary muon-spin 
rotation ($\mu$SR) experiments on \ZVB\ (see Fig.~\ref{fig:ZrVB2_TF_MuSR}).  
Once implanted in matter, spin-polarized muons act as microscopic probes of magnetism, 
which upon decay emit positrons preferentially along the muon-spin direction. 
From the spatial anisotropy of the emitted positrons (i.e., the asymmetry signal) 
one can reveal the distribution of the local magnetic fields. \cite{Blundell1999,Yaouanc2011}
In our case, by applying of 0.6\,T, a regular flux-line lattice (FLL) 
develops in the superconducting phase below $T_c$. By uniformly sampling the FLL muons experience 
an additional relaxation $\sigma_{\mathrm{sc}}$, which is related to the absolute magnetic 
penetration depth $\lambda$ via:\cite{Brandt1988,Brandt2003}

\begin{equation}
\label{eq:lambda}
\frac{\sigma_{sc}^2}{\gamma^2_{\mu}}=0.00371 \cdot \frac{\varPhi_0^2}{\lambda^4}.
\end{equation}

Here $\varPhi_0 =2.068\times10^{-3}$\,T$\mu$m$^2$ is the magnetic flux quantum and 
$\gamma_{\mu} = 2\pi \times 135.53$\,MHz/T, the muon gyromagnetic ratio. 

Figure~\ref{fig:ZrVB2_TF_MuSR} shows the temperature dependence of $\sigma_{sc}$, proportional to the superfluid density ($\sigma_{sc} \propto n_s \propto \lambda^{-2}(T)$), together with a numerical fit with an average-field model $1/\lambda^2(T)=(1/\lambda^2(0))[1-(T/T_c)^n]$, which gives $1/\lambda^2(0)=1.43\pm0.2$\,$\mu$m$^{-2}$ and $n=2.0\pm0.1$. Subsequently, by using the relation $\lambda_{ab}(0)= \lambda_{\mathrm{eff}}(0)/1.31$, we estimate the in-plane magnetic penetration depth $\lambda_{ab}(0) = 638\pm11$\,nm.  
This value is close to the one determined via macroscopic methods (see above), but it is very different from 
$\lambda_{ab}(0) = 100$\,nm, also measured via $\mu$SR in MgB$_2$.\cite{Niedermayer2002} 
This difference can be accounted for by considering the rather small electronic density of states in \ZVB\ compared to that 
in MgB$_2$, compatible with the very different NMR relaxation-rate values reported in 
Table~\ref{tab:summary}.

As a final note, we recall that the choice of the applied transverse field is crucial for the correct 
determination of the field penetration depth. Indeed, detailed studies of the magnetic field dependence of 
the muon-spin depolarization rate in MgB$_2$ (see, e.g., Figs.~1 and 2 in Ref.\onlinecite{Niedermayer2002}) 
have shown strong pinning effects in low applied fields (below 0.3\,T). These imply a considerably distorted 
vortex lattice leading to a strong decay of the muon asymmetry and hence underestimated $\lambda_{ab}(0)$ 
values. However, since in fields exceeding 0.3\,T, only weak or no pinning effects were observed, we are confident
that by applying a transverse field of 0.6\,T, our results reflect the penetration depth. 

\vspace{-5pt}
\section{Summary and conclusions\label{sec:Conclusions}}

From SQUID magnetometry, NMR, resistivity, and preliminary $\mu$SR experiment data on \ZB, \HB, \ZVB,
and \HVB, we argue that the latter two samples are $s$-wave superconductors, resulting from electron doping via
$d$ orbitals of vanadium. On the other hand, in MgB$_2$ the peculiar self-doping 
exists thanks to the boron-like electrons at the Fermi level; at the same time also the $s$ states
of Mg donate their electrons to the boron-derived conduction bands (\textit{metallic} B).
It is also worth mentioning that B vacancies (a common defect occurring in diborides) enhance
the DOS even further (more than twice for doping of about 0.5\,\%).\cite{Dahlqvist2015} Besides the many
similarities with MgB$_2$, the lower $T_c$ values (7.31 and 8.33\,K instead of 39\,K) are accounted for
by the differences in the electronic structure and Fermi surface. The DOS per unit cell of MgB$_2$ at the Fermi level
(0.719\,states/eV) is dominated (about 60\%)  by the $p$ orbitals of B atoms, while in the case of
the undoped transition-metal diborides (\ZB, for instance) 80\% of the DOS (0.130\,states/eV)\cite{Shein2008}
derives from the $d$ orbitals of the cation. Furthermore, our data support the hypothesis that we are dealing
with new two-band two-gap $s$-wave superconductors, since the coexistence of a normal and a superconducting
phase below $T_c$ suggests an anisotropy of the upper critical field, as well documented in the paradigmatic case of
MgB$_2$. More direct evidence for this assumption could be obtained by various experiments probing single 
 crystals of \ZVB\ and \HVB.

\begin{acknowledgments}
This work was financially supported in part by the Schweizerische Nationalfonds zur F\"{o}rderung der Wissenschaftlichen
Forschung (SNF).
\end{acknowledgments}


%

\end{document}